\documentclass[a4paper,11pt]{article}
\usepackage{graphicx}
\usepackage{filecontents}
\usepackage{amssymb}

\usepackage{times}
\usepackage{mathptm}
\usepackage[bf]{caption}

\title{A reproduction of the Milky Way’s Faraday rotation measure map in galaxy simulations from global to local scales}

\author{Stefan Reissl$^{1}$\thanks{E-mail: reissl@uni-heidelberg.de},\
  \ Ralf S. Klessen$^{1,2}$, Eric W. Pellegrini$^{3}$,\\ Daniel
  Rahner$^{1}$, R{\"u}diger Pakmor$^{3}$, Robert Grand$^{4}$, \\ Facundo G{\'o}mez$^{5,6}$, Federico Marinacci$^{7}$, and Volker Springel$^{3}$}

\begin{document}

\maketitle
{\small\quad\\$^{1}$Universit{\"a}t Heidelberg, Zentrum f\"{u}r Astronomie, Institut f\"{u}r Theoretische Astrophysik, Albert-Ueberle-Str. 2, 69120 Heidelberg, Germany\\
$^{2}$Universit{\"a}t Heidelberg, Interdisziplin{\"a}res Zentrum f{\"u}r Wissenschaftliches Rechnen,Im Neuenheimer Feld 205, 69120 Heidelberg, Germany\\
$^{3}$Max-Planck-Institut f{\"u}r Astrophysik,
Karl-Schwarzschild-Str. 1, 85748 Garching, Germany\\
$^{4}$Astrophysics Research Institute, Liverpool John Moores University, 146 Brownlow Hill, Liverpool, L3 5RF, UK\\
$^{5}$Instituto de Investigaci\'on Multidisciplinar en Ciencia y Tecnolog\'ia, Universidad de La Serena, Ra\'ul Bitr\'an 1305, La Serena, Chile\\
$^{6}$Departamento de F\'isica y Astronom\'ia, Universidad de La Serena, Av.
Juan Cisternas 1200 Norte, La Serena, Chile\\
$^{7}$Institute for Theory and Computation, Harvard-Smithsonian Center
for Astrophysics, 60 Garden Street, Cambridge, MA 02138, USA}

\normalsize
\vspace{1.0cm}
{\bf Magnetic fields are of critical importance for our understanding
  of the origin and long-term evolution of the Milky Way. This is due to their decisive role in the dynamical evolution of the interstellar medium (ISM) and their influence on the star-formation process
  \cite{McKey2007,Ferriere2001,Heiles2012,Beck2015}.  Faraday rotation measures
  (RM) along many different sightlines across the Galaxy are a primary
  means to infer the magnetic field topology and strength from
  observations \cite{Morris1964,Oppermann2012,Sun2015,Hutschenreuter2022}.  However, the
  interpretation of the data has been hampered by the failure of
  previous attempts to explain the observations in theoretical models
  and to synthesize a realistic multi-scale all-sky RM map
  \cite{Beck2016,Butsky2017,Pakmor2018}.  We
  here utilize a cosmological magnetohydrodynamic (MHD) simulation of
  the formation of the Milky Way, augment it with a novel star cluster
  population synthesis model for a more realistic structure of the
  local interstellar medium \cite{Rahner2017,Pellegrini2020-POP}, and perform detailed polarized radiative transfer calculations on the resulting model
  \cite{Brauer2017, Reissl2018X}.  This yields a faithful first
  principles prediction of the Faraday sky as observed on Earth.  The
  results reproduce the observations of the Galaxy not only on global
  scales, but also on local scales of individual star-forming clouds.
  They also imply that the Local Bubble \cite{Alves2018,Zucker2022} containing
  our Sun dominates the RM signal over large regions of the sky.
  Modern cosmological MHD simulations of the Milky Way's formation,
  combined with a simple and plausible model for the 
  fraction of free electrons in the ISM, explain the RM
  observations remarkably well, thus indicating the emergence of a
  firm theoretical understanding of the genesis of magnetic fields in
  our Universe across cosmic time.\\}

Magnetic fields significantly influence the kinematical and morphlogical properties of the ISM and contribute to regulating the birth of new generations of stars \cite{Klessen2016}. To better understand this connection, several observational techniques have been developed and perfected over the last century. For example, dust polarization measurements of aligned dust grains \cite{Andersson2015,Planck2016XXXV} and synchrotron emission \cite{Beck2015,Lazarian2017,Reissl2019} allow us to infer the projected line-of-sight (LOS) field orientation, while estimates of the LOS field strength can be obtained from the Zeeman effect \cite{Crutcher1999}. However, these methods are only applicable to a limited set of parameters \cite{Brauer2017}, and additional uncertainties arise from our incomplete understanding of the microphysics involved such as grain alignment mechanisms \cite{Andersson2015}. A complementary approach is to determine the  characteristic Faraday rotation measure (RM). It is based on the fact that polarized radiation can change its polarization angle as it passes though a magnetized and ionized medium. The observed signal depends on the magnetic field strength and direction as well as on the density of free electrons, and on the radiation frequency \cite{Rybicki1979}. 

Since the early 1960's, numerous attempts were made to reconstruct an all-sky RM map of the Milky Way \cite{Morris1964,Oppermann2012,Sun2015} from observations of pulsars and extra-galactic background sources. Complementary synthetic data has proven its worth for systematically interpreting and analyzing this plethora of observations \cite{Beck2016,Butsky2017,Pakmor2018}. However, all existing approaches are hampered by the fact that the distribution of thermal electrons as well as the detailed structure of the magnetic field are not well known. While the large-scale properties are usually well constrained \cite{Pakmor2014,Hennebelle2018}, crucial information about individual star-forming regions and clouds is missing, which is needed to reproduce the observed small-scale features.   

Our approach goes beyond the current state-of-the-art and employs data from a high-resolution cosmological simulation to reconstruct the large-scale properties of the galaxy combined with a novel star cluster population synthesis model, which introduces the missing small-scale physics. Specifically, we take the Au-6 galaxy from the Auriga project \cite{Grand2017}, which is able to reproduce the global star-formation rate and structure of the Milky Way very well, while at the same time predicting the amplification of minute primordial magnetic seed fields to micro-Gauss strength over secular timescales.  We keep the overall gas density and magnetic field structure, but we discard the original stellar population and distribution of free electrons. We then synthesize a new population of star clusters and calculate the corresponding radiative and mechanical feedback based on the WARPFIELD cloud-cluster evolution method \cite{Rahner2017, Rahner2019}. Next, we obtain the corresponding emission from each cluster across the electromagnetic spectrum  \cite{Pellegrini2020-EMP}, use the polarized radiative transfer code {\sc POLARIS} \cite{Reissl2016} to build up the spatially varying interstellar radiation field in the galaxy, and from that reconstruct the distribution of free electrons in the ISM, as discussed by Pellegrini and colleagues \cite{Pellegrini2020-POP}. Finally, we perform a second {\sc POLARIS} sweep to calculate synthetic Faraday RM all-sky maps \cite{Reissl2019} from varying positions within the galaxy.

The total integrated angle of linear polarization for an observer follows as
\begin{equation}
\chi_{\rm obs}=\chi_{\rm source}+RM\times \left(\frac{c}{\nu}\right)^2 \, ,
\end{equation}
where $\chi_{\rm source}$ is the polarization angle of the source and $\nu$ is the observed frequency. The integral
\begin{equation}
RM = \frac{1}{2\pi}\frac{e^2}{m_{\rm e}^2 c^4} \int_{0}^{s_{\mathrm{obs}}} n_{\rm th}(s) B_{||}(s) \mathrm{d}s
\label{eq:RMDefinition}
\end{equation}
defines the rotation measure \cite{Burn1966} for the non-relativistic limit along the LOS $s$ towards the observer, $c$ is the speed of light, $e$ and $m_{\rm e}$ are the electron charge and mass, respectively, and $B_{||}$ is the LOS magnetic field component. Altogether we compute high-resolution Faraday RM all-sky maps with a total number of  $N=786,432$ pixels for ten distinct observer positions within the Au-6 galaxy (which we denote P01 - P10). They are placed at roughly the same distance to the galactic center as the Sun (that is at galactocentric radii $8\,\mathrm{kpc} \le R \le 10\,\mathrm{kpc}$) and are located in a gas density cavity similar to the Local Bubble that defines our own Galactic environment \cite{Alves2018, Zucker2022}, as illustrated in Figure 6 of Pellegrini and colleagues \cite{Pellegrini2020-POP}.. 


\begin{figure*}
\begin{center}
  \begin{minipage}[c]{1.0\linewidth}
    \begin{center}
     \includegraphics[width=1.02\textwidth]{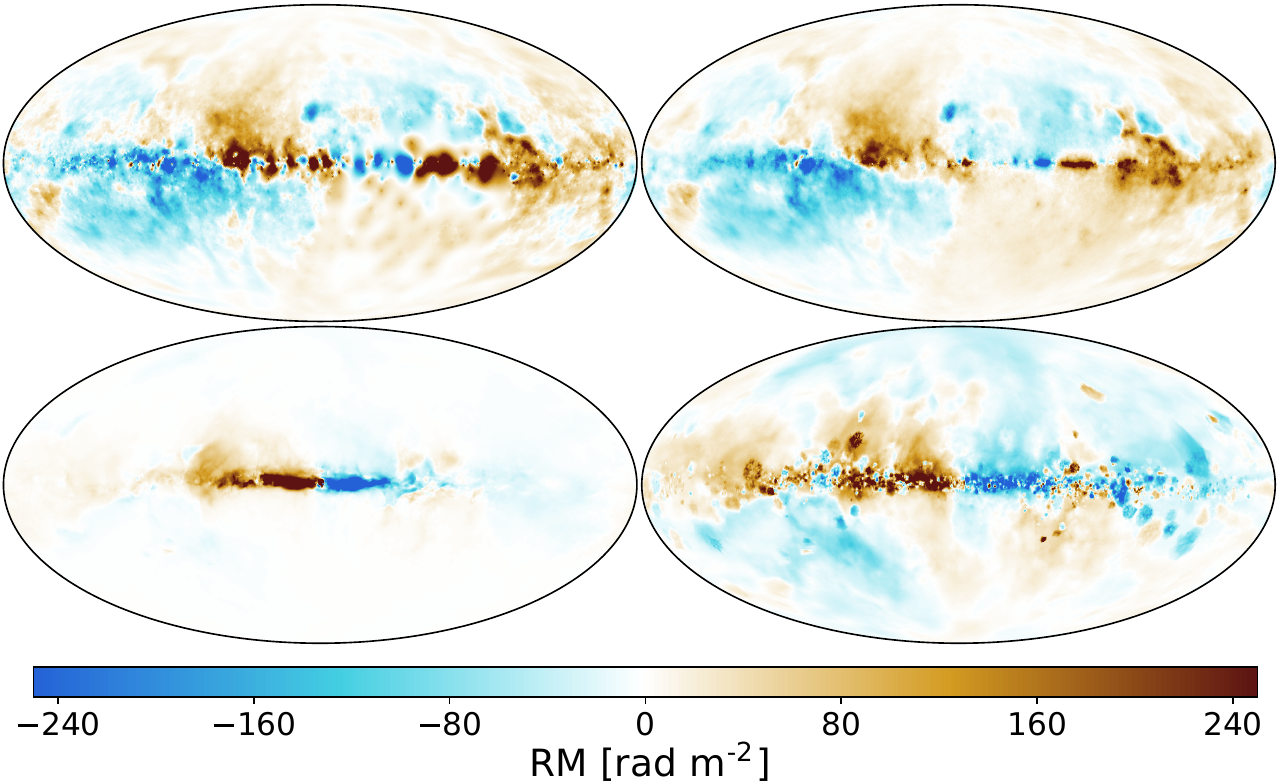}
     \vspace{-0.5mm}
     \caption{\emph{All-sky Faraday RM map of the Milky Way and of a model galaxy from the Auriga suite of cosmological simulations.} In the top row we show the observed maps of O12 (left) and the data from H22 (right). In the bottom row (left) we provide the synthetic RM map constructed from the original Au-6 data \cite{Pakmor2018}. The lack of signal at large galactic latitudes and the failure to reproduce small-scale features is evident. This problem is remedied by combining the galaxy with a realistic small-scale model of star-cluster formation and evolution and a self-consistent reconstruction of the density of free electrons \cite{Pellegrini2020-POP}, while keeping the original magnetic field structure of the cosmological simulation \cite{Pakmor2017}. The resulting  resulting RM map is shown at the bottom right and very similar to the Milky Way observations. }
     \label{fig:RMAllSkyMaps}
     \end{center}
  \end{minipage}
\end{center}
\end{figure*}

\begin{figure*}
  \begin{minipage}[c]{1.0\linewidth}
	    \includegraphics[width=0.51\textwidth]{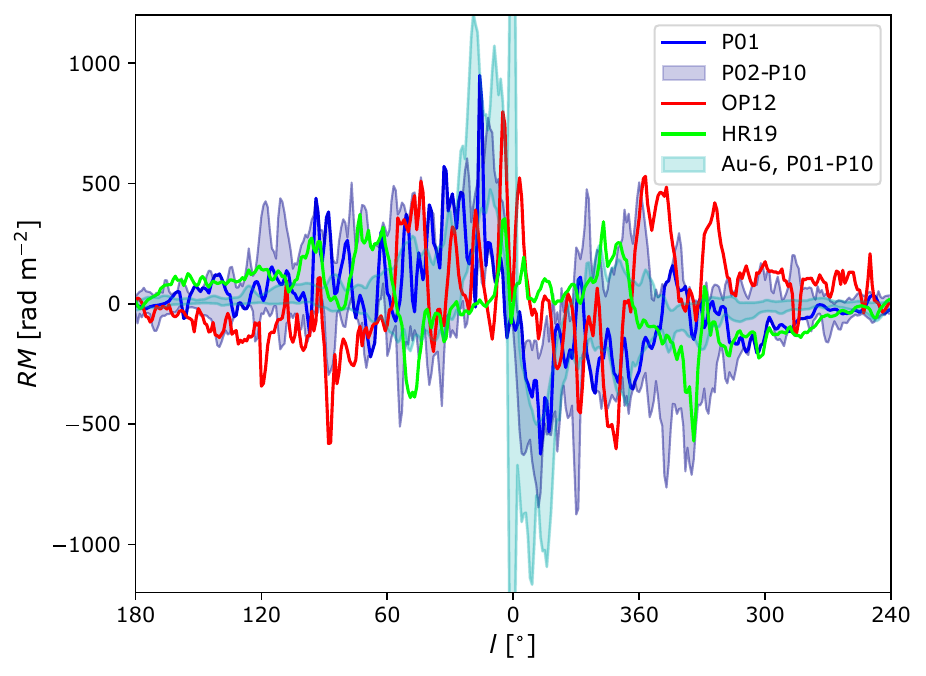}
	    \includegraphics[width=0.49\textwidth]{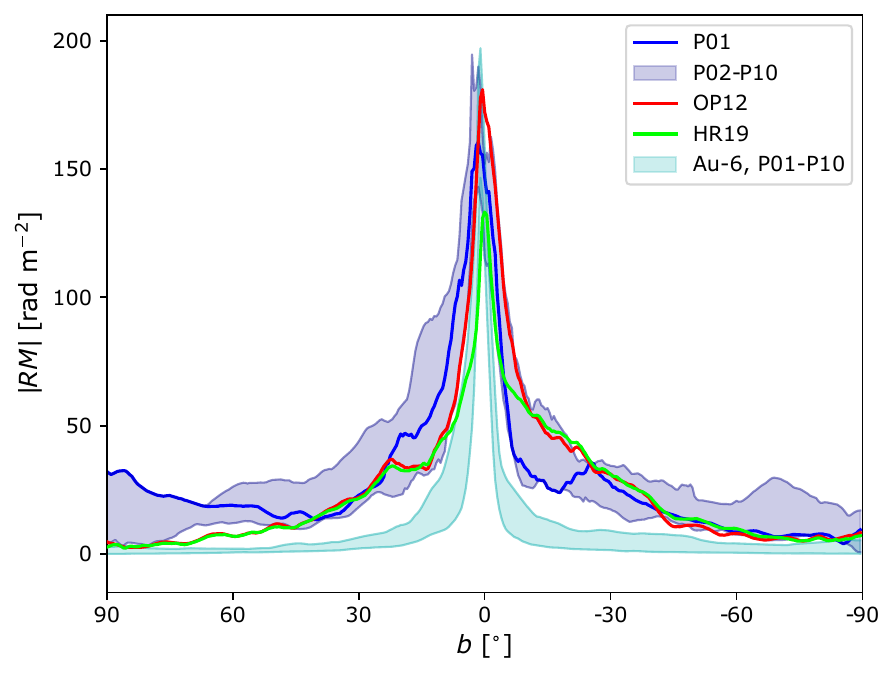}
	    \vspace{-3mm}
	    \caption{ \emph{Average magnitude of the Faraday RM along
                the Galactic longitude $l$ (left panel) and Galactic
                latitude $b$ (right panel).}  The magnitude of the Faraday
              RM has been averaged over latitude within $|b|<1.2^\circ$
              when shown as a function of longitude for all of our
              all-sky RM maps. The amount of RM differs for each of
              the observer positions (synthetic and our own within the
              Milky Way) as a result of the local electron
              distributions and magnetic field directions,
              respectively. However, all RM profiles show the same
              trend from positive to negative values at $l=0^\circ$ as
              the large scale toriodal field component reverses its
              direction with respect to the observer.  Also, the
              general synthetic and observed trends agree well with a
              peak at $b=0^\circ$ and a minimum towards the poles.}
     \label{fig:RMProfile}
  \end{minipage}
\end{figure*}

\begin{figure*}
  \begin{minipage}[c]{1.0\linewidth}
	    
	    \includegraphics[width=0.51\textwidth]{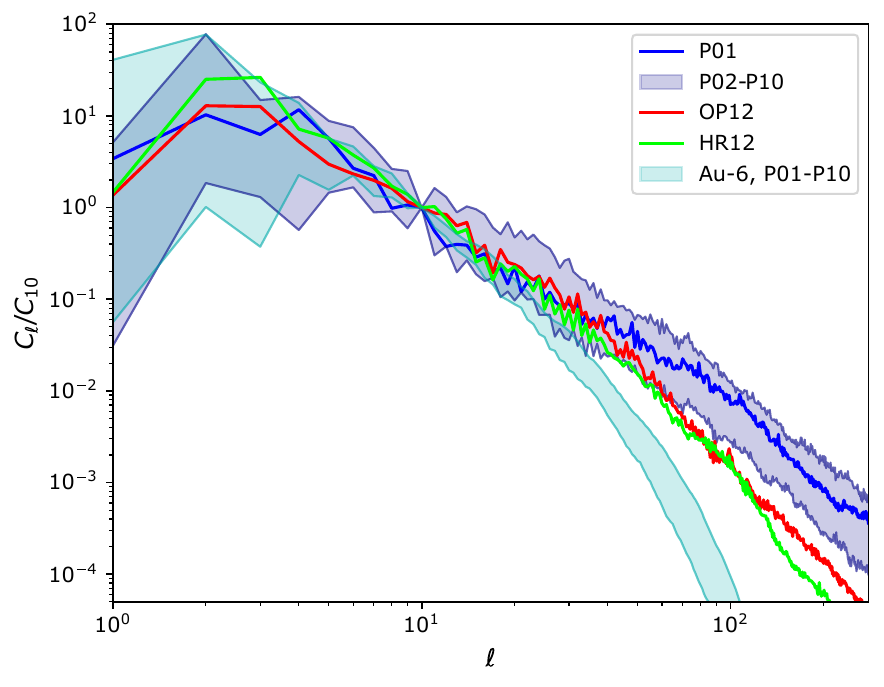}
	    \includegraphics[width=0.51\textwidth]{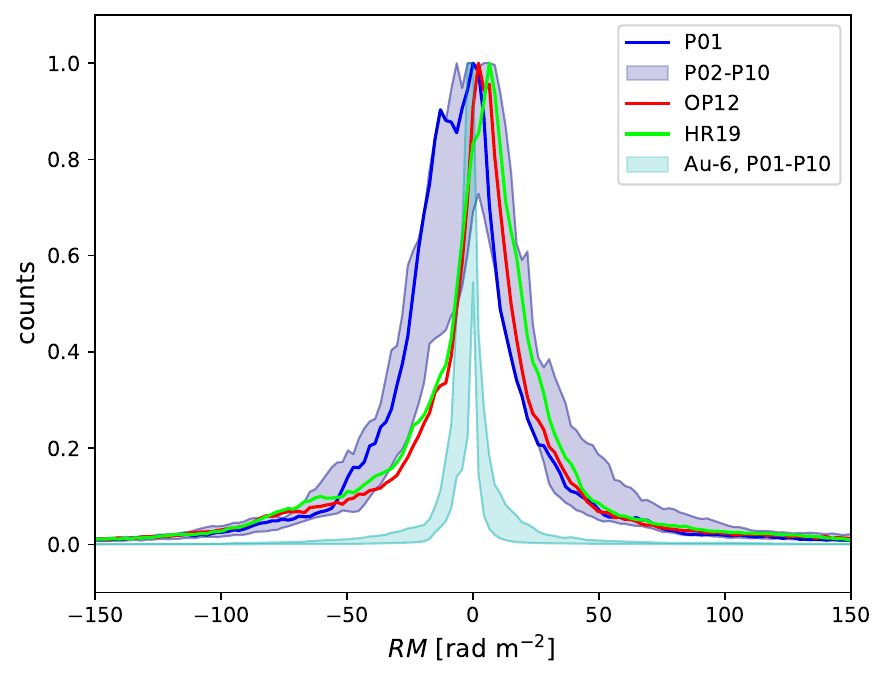}
\vspace{-3mm}
\caption{ \emph{Multipole spectrum of RM maps for ten different
    observer positions compared to observational data for the Milky
    Way.}  Left panel: Direct multipole spectrum as a
  function of multipole moment $\ell$ for ten different observer
  positions (P01-P10) within the origin Au-6 galaxy and our modified galaxy version and compares
  with the reconstructed RM observational data of O12 and H22. A multipole moment of
  $\ell=100$ corresponds roughly to a resolution of
  $14.4^\circ$. At higher $\ell$ the synthetic map of the modified galaxy 
  shows more structure all over the sky in comparison with the reconstructed O12 and H22 data. All spectra are normalized at $\ell=10$ for better comparison. Right panel: Histogram of the RM healpix maps shown in Fig. \ref{fig:RMAllSkyMaps}. All distributions are normalized by their peak values.}
     \label{fig:RMSpectrum}
  \end{minipage}
\end{figure*}

The existing RM all-sky maps of the Milky Way are built from an ensemble of extragalactic polarized radio sources. Certain patches of the sky with missing data or with reduced coverage are reconstructed applying Bayesian statistics, and consequently, these regions appear to be smoother than the average. We compare our results with the observed all-sky RM maps presented by Oppermann and collaborators \cite{Oppermann2012} and by  Hutschenreuter and colleagues \cite{Hutschenreuter2022}, who included a larger source number and employed a more sophisticated reconstruction algorithm (O12 and H22 hereafter). We note that it has recently been pointed out \cite{Shanahan2019} that high RM sightlines may have gone undetected for decades, either from instrumental limitations or from biases in the source selection, and that this introduces a systematic shift of these maps towards low RM values. 
 
Figure~\ref{fig:RMAllSkyMaps} shows the O12 and H22 data {\em (top row)} as well as two synthetic RM maps from the Auriga galaxy {\em (bottom row)} from one exemplary observers position (P01) at the solar circle. The synthetic map at the left takes the galaxy as is \cite{Pakmor2018}, based on the data of the cosmological simulation only, and the right one combines the simulation data with the detailed model for star-cluster formation and evolution   \cite{Pellegrini2020-POP} described above. 
It is immediately obvious that the Auriga-only maps misses small-scale features, whereas the full model agrees remarkably well the observations. It has a comparable level of fluctuations on small angular scales and gives the right RM magnitude in all regions of the sky. We also note that all maps exhibit a reversal of the magnetic field direction at the Galactic center, as indicated by a transition from positive to negative RM values. This transition is a distinct feature of the magnetic field morphology of the Milky Way,  revealing a globally toroidal field structure. It is also well visible in  Figure~\ref{fig:RMProfile}, where we plot the average RM along the Galactic longitude $l$ {\em (left)} and latitude $b$ {\em (right)} coordinates for all observer positions in the model galaxy. The field reversal in the center is a global property of the disk and therefore independent of the location of the measurement. Similarly, all models exhibit the highest amount of RM near the disk midplane ($b=0^\circ$) with decreasing values towards the poles ($b=\pm 90^\circ$), again in agreement with O12 and H22, indicating that the magnetic field quickly decreases further up and down into the Galactic halo. We also mention that our model slightly overestimates the amount of RM at large $b$ with an offset of $0-30\ \mathrm{rad\ m}^{-2}$. Since the signal at large latitudes $b$ is mostly produced by material that is nearby \cite{Pakmor2018}, this deviation emphasizes again the importance of properly accounting for the immediate surrounding of the Sun, and it is an indication of the limitations of current galaxy formation simulations when it comes to reproducing the Local Bubble \cite{Alves2018, Zucker2022}.

For a more quantitative comparison, we compute the multipole expansion of the RM
maps and display the result in Figure \ref{fig:RMSpectrum}. Our
synthetic RM spectra are consistent with the O12 and H22 data for all
multipole moments $\ell \lesssim 30$ at all observer positions, whereas we are predicting more small-scale structures at larger $\ell$. We speculate that this may be a result of the bias in the Bayesian statistics of O12 and H22 towards lower RM values \cite{Shanahan2019}. In contrast, the expansion of the original AU-6 galaxy \cite{Pakmor2018} deviates from the observations already above $\ell \approx 10$ indicating again importance of small-scale physics for the interpretation of the Milky Way data. We stress that including a realistic star-cluster population synthesis model is indispensable for reproducing the observed small-scale (large $\ell$) features. We also note that similar conclusions have been reached by Beck and collaborators \cite{Beck2016}, however, they did so by introducing a small-scale random magnetic field component still lacking the contribution of a proper multi-scale thermal electron model.

Altogether, we find that the general trends in the RM maps presented here agree well for different positions within the galaxy, and they all reproduce the observed data. Consequently, we conclude that current high-resolution MHD simulations of formation and evolution of the Milky Way in a cosmological context, combined with adequate models of star formation and stellar feedback, can well explain the properties of magnetic fields in spiral galaxies. This marks an important step forward in our theoretical understanding of magnetic field amplification by various forms of the dynamo process acting in these systems \cite{Pakmor2018,Rieder2016}.  We note that the methods presented here can also be applied to RM observations in the high-redshift Universe \cite{Mao2017} and can thus help to monitor the genesis of magnetic fields over cosmic timescales. However, we also caution that most of the RM signal above and below the Galactic plane might be dominated by the local environment \cite{Pakmor2018}. Subsequently, maps of the Faraday rotation of the Milky Way cannot be adequately interpreted without knowledge of the conditions in our Local Bubble \cite{Alves2018,Zucker2022}. This has been proposed before \cite{Sun2015, Pakmor2018}, and it is also implied by the dust polarization measurements of the Planck satellite \cite{Alves2018}. Distinct observers in different parts of the Galaxy would see different local magnetic field configurations and electron densities. Our results suggest that current measurements of the Milky Way RM carry a level of uncertainty that was previously not fully appreciated and that can only be accounted for on a statistical basis by detailed modeling efforts as presented here.

\section*{Methods}
To construct the synthetic all-sky Faraday RM map of the Milky Way, we
take one of the simulations from the Auriga project
\cite{Grand2017}. The galaxy Au-6 is the result of a very
high-resolution cosmological MHD zoom-in simulation from initial
conditions that are specifically selected to reproduce key features of
the Local Group. The calculation includes line cooling, stellar
evolution, galactic winds, and the growth of black holes and their associated
AGN feedback \cite{Grand2017}. All simulations include
self-consistently evolving magnetic fields on a Voronoi grid
\cite{Pakmor2014} with the moving mesh code {\sc AREPO}
\cite{Springel2010}. The galaxy Au-6 is selected as Milky Way analogue based on its size, total mass, and star formation rate. 

In the next step we follow the procedure as outlined by Pellegrini and colleagues \cite{Pellegrini2020-POP} (hereafter P21) and synthesize a new cluster population and electron fractions that are more faithful to star-forming physics and small scale density structures know from observations. To do so, we replace the original star particles and electron fractions from the simulation and instead reconstruct this information from first principles employing the WARPFIELD cloud/cluster evolution model \cite{Rahner2017, Rahner2019}. It follows the time evolution and structure of the stellar wind bubble, H{\sc II} region, and protodissociation region (PDR) surrounding a cluster of massive stars in spherical symmetry. WARPFIELD accounts self-consistently for the physics of stellar winds, supernovae, radiation pressure, ionization, and gravity. It solves explicitly for the density structure adopted by the gas in response to the action of these various feedback processes, and therefore allows one to account for the evolution of the luminosity and emerging spectrum. 

We determine the local gas mass within equidistant annuli of the Au-6 galaxy, using linearly spaced radial bins originating at the galactic center-of-mass and sample the cluster mass function by randomly depositing mass with a rate obtained from the Kennicutt-Schmidt \cite{kennicutt98b} relation. Per annulus and cluster a random location is selected. If the total gas mass within a distance less than $50\ \mathrm{pc}$ from that location is larger than the cluster mass we have drawn, then the cluster is placed there, and the corresponding gas mass is subtracted from the original Voronoi grid of the Au-6 galaxy. As a result of this procedure, the star clusters are typically inserted near density peaks. This is consistent with observations, where young clusters are seen in the vicinity of dense molecular gas, but no longer are deeply embedded into their parental clouds due to efficient stellar feedback \cite{Krumholz2019}. This leads to a realistic stellar population that is characterized by the number of clusters for a given mass and age in each annulus. An illustratino of this approach is presented in Figure 6 of P21. 

The physical properties of each cluster including all emission properties are obtained from the WARPFIELD database \cite{Rahner2017, Rahner2019} combined with the spectral synthesis code {\sc CLOUDY}\footnote{http://www.nublado.org/} $\rm v17.00$ \cite{Ferland2017}. This information \cite{Pellegrini2020-EMP} is then used to compute the impact of ionizing radiation from each individual cluster on the ambient Galactic ISM in order to (re)populate the entire Au-6 Voronoi grid with thermal electrons. We assume solar metallicity, and cosmic ray ionization rates and an interstellar radiation field (ISRF) that is representative of the Milky Way \cite{Klessen2016}. Note that we neglect the population of old field stars in our analysis, because they do not contribute to the ISRF at ionizing frequencies. The resulting distribution of free electrons agrees very well with the observationally-inspired models for the Milky Way by Cordes \& Lazio \cite{Cordes2002} and by Yao et al. \cite{Yao2017}. This is illustrated in Figure 8 of P21. For a detailed description of the magnetic field structure of the Auriga-6 galaxy, we refer to Pakmor et al. \cite{Pakmor2017}. With this approach, we have all necessary information to compute the integral (\ref{eq:RMDefinition}) through arbitrary sightlines through the galaxy. 

Finally, we employ the radiative transfer (RT) code {\sc POLARIS}\footnote{http://www1.astrophysik.uni-kiel.de/${\sim}$polaris/} \cite{Reissl2016} capable of dust polarization calculations \cite{Reissl2018X} as well as Zeeman splitting line RT \cite {Brauer2017,Reissl2018X} and solve the RT problem on the native Voronoi grid of {\sc AREPO} simulations. We calculate the resulting synthetic Faraday RM all-sky maps from ten selected positions in the Au-6 galaxy integrating equation (\ref{eq:RMDefinition}) for many different sightlines across the entire galaxy and covering the entire sky with high angular resolution \cite{Reissl2019}. For our calculations of the synthetic RM maps we apply a HEALPIX\footnote{http://healpix.jpl.nasa.gov} resolution of $N_{\mathrm{side}}=256$ leading to a total amount of $786\ 432$ pixel. This resolution is identical to the observed H22 map \cite{Hutschenreuter2020} but larger than O12 map by a factor of four. The location of the fictitious observers in the Au-6 galaxy are selected such that the distance to the galactic center is between $8\,$kpc and $10\,$kpc, with the solar values being 8.5$\,$kpc, and that they are placed in a Local Bubble like region, where previous supernovae have created a low-density cavity, as illustrated in Figure 6 of P21.

\section*{Code Availability }
Cluster properties and ionizaiton are calulated with the WARPFIELD code \cite{Rahner2017, Rahner2019} and the spectral synthesis code {\sc CLOUDY} $\rm v17.00$ (http://www.nublado.org/), respectively. Cosmological simulations are performed by the moving mesh code {\sc AREPO} \cite{Springel2010} ({https://arepo-code.org/wp-content/userguide/index.html}) and the RT post-processing we utilize the RT code {\sc POLARIS} \cite{Reissl2016} ({https://portia.astrophysik.uni-kiel.de/polaris/}). We used Python and its associated libraries including {\sc astropy}, {\sc numpy}, and {\sc matplotlib} for data analysis and presentation. 

\section*{Corresponding Author}
The corresponding author is Stefan Reissl. Please send any requests for further information or data to {\tt reissl@uni-heidelberg.de}.

\section*{Acknowledgements}
S.R., R.S.K., E.W.P., and D.R. acknowledge  support  from  the  Deutsche  Forschungsgemeinschaft in the Collaborative Research Center (SFB 881, ID 138713538)  ``The Milky Way System'' (subprojects A1, B1, B2, and B8) and from the Heidelberg Cluster of Excellence (EXC 2181, ID 390900948) ``STRUCTURES: A unifying approach to emergent phenomena in the physical world, mathematics, and complex data'', funded by the German Excellence Strategy. R.S.K. also thanks for funding form the European Research Council in the ERC Synergy Grant ``ECOGAL -- Understanding our Galactic ecosystem: From the disk of the Milky Way to the formation sites of stars and planets'' (ID 855130).  RG acknowledges support from an STFC Ernest Rutherford Fellowship (ST/W003643/1).” F.A.G. acknowledges financial support from CONICYT through the project FONDECYT Regular Nr. 1181264, and funding from the Max Planck Society through a Partner Group grant. The project benefited from computing resources provided by the State of Baden-W\"urttemberg through bwHPC and DFG through grant INST 35/1134-1 FUGG, and from the data storage facility SDS@hd supported through grant INST 35/1314-1 FUGG. The Heidelberg team also thank for computing time provided by the Leibniz Computing Center (LRZ) for project pr74nu.
 
\section*{Author Contributions}
S.R. has run all polarized radiative transfer calculations and has performed most of the analysis. The text was jointly written by S.R. and R.S.K. The WARPFIELD cloud-cluster evolution model was mostly contributed by E.W.P. and D.R. The Augiga-6 data and support with the data handling have been provided by R.P., R.G., F.G., F.M., and V.S.

\bibliographystyle{plain}
\bibliography{shortbib}
\end{document}